# Formation of magnetic glass in calcium doped $YBaCo_2O_{5.5}$ cobaltites

Tapati Sarkar*, V. Pralong and B. Raveau

*Laboratoire CRISMAT, UMR 6508 CNRS ENSICAEN,*

*6 bd Maréchal Juin, 14050 CAEN, France*

**Abstract**

The d.c. magnetization and magnetic relaxation studies of the calcium doped samples, $Y_{0.95}Ca_{0.05}BaCo_2O_{5.5}$ and $YBa_{0.95}Ca_{0.05}Co_2O_{5.5}$, show the existence of a magnetic glass like behaviour in the family of cobaltites for the first time. Our investigations reveal glass-like arrest of kinetics at low temperature which prevents the system from reaching its magnetic ground state. We show that the low temperature state of these calcium doped phases, which consists of coexisting antiferromagnetic and ferro (or ferri) magnetic phase fractions, can be tuned in a number of ways. Our observations establish that the low temperature state of this oxide is not in thermal equilibrium. The glassy state is formed with the assistance of an external magnetic field, which makes it distinctly different from the more well known metastable state, the spin glass state. The cooling field can tune the fractions of the coexisting phases, and the glass-like state formed at low temperature can also be devitrified by warming the sample. The role of Ca doping in the appearance of these phenomena is discussed in terms of phase separation, involving $Co^{3+}$ disproportionation into $Co^{4+}$ ferromagnetic clusters and $Co^{2+}$ antiferromagnetic clusters.

* Corresponding author: Tapati Sarkar
  e-mail: tapati.sarkar@ensicaen.fr

**Introduction**

Cobaltites have gained increased prominence in the scientific community during the last decade, due to the rich variety of phenomena they exhibit. Among the cobaltites, the "112" ordered oxygen deficient perovskites $LnBaCo_2O_{5.5+\delta}$ discovered more than fifteen years ago [1, 2], whose magnetoresistance properties [3] were shown to be remarkable, are currently being studied by many authors, showing that their physical properties are strongly influenced by various phenomena such as charge ordering, oxygen non-stoichiometry, order-disorder phenomena about anionic vacancies, as well as phase separation phenomena and possible spin transitions [4 – 24].

One important characteristic of these "112" cobaltites deals with the existence of multiple magnetic transitions in a wide temperature range i.e., a paramagnetic to ferromagnetic transition around room temperature followed by a ferromagnetic to antiferromagnetic transition at lower temperature, which have been the subject of numerous investigations [4 – 24]. Among the unusual features which have been observed in these oxides, the presence of significant thermomagnetic irreversibility at low temperature in $LaBaCo_2O_{5.5}$, manifested by a large divergence between the zero field cooled (ZFC) and field cooled (FC) M(T) curves, which persists even at high magnetic fields [23], is of interest. The authors have indeed shown that the magnetization drop below $T_N$ was suppressed in the presence of high magnetic fields, indicating a phase competition between the ferromagnetic and antiferromagnetic phases. The antiferromagnetic phase was seen to have nonzero value of magnetization, indicating the persistence of ferromagnetic-like interactions below $T_N$. Investigation of the magnetotransport properties of this compound [24] revealed that external magnetic fields readily induce a magnetoresistance (MR) effect, thereby indicating the subtle balance between the ferromagnetic and antiferromagnetic phases. It was seen that the maximum MR was obtained near the region of the ferromagnetic – antiferromagnetic phase boundary (~ 245 K).

Though it is not limited to the La-phase, the problem of phase separation and its impact upon magnetism in the $LnBaCo_2O_{5.5}$ phases has been little studied at low temperature i.e., below $T_N$. In this respect, the recent studies of the calcium doping of the "112" cobaltite $YBaCo_2O_{5.5}$ [20, 21] bring out an important contribution to the knowledge of magnetism at low temperature in these systems. They show that, whatever be the doping site i.e., yttrium for $Y_{0.95}Ca_{0.05}BaCo_2O_{5.5}$ [20] or barium for $YBa_{0.95}Ca_{0.05}Co_2O_{5.5}$ [21], the calcium doping induces a large expansion of the ferromagnetic state to the detriment of the long range antiferromagnetic order. In both the cases, the authors have suggested the existence of a two-phase scenario. Bearing in mind these results, we have revisited the magnetic properties of these two Ca-doped phases, focusing our attention on their lower temperature behaviour i.e., below 150 K, using detailed d.c. magnetometry measurements. We show that whatever be the doping site, Ba or Y, the "112" phases $YBa_{0.95}Ca_{0.05}Co_2O_{5.5}$ and $Y_{0.95}Ca_{0.05}BaCo_2O_{5.5}$, exhibit, at low temperature, a novel metastable magnetic state, which was not seen before in the "112" family. This state corresponds to a nonequilibrium glassy behaviour, which arises from a kinetic arrest of the ferromagnetic to antiferromagnetic phase transition, i.e., results from the viscous retardation of growth of the low temperature antiferromagnetic phase out of the supercooled ferro (or ferri) magnetic phase. In fact, these oxides fall in the category of the recently identified *magnetic glasses* [25 – 28], which consists of ferro (or ferri) magnetic and antiferromagnetic clusters frozen randomly in experimental time scale, formed with a dynamics similar to that of structural glasses. Importantly, the formation of this glassy state in these oxides requires the assistance of an external magnetic field, and hence, is distinctly different from the more well-known spin glass state.

**Experimental**

Polycrystalline samples of $YBa_{0.95}Ca_{0.05}Co_2O_{5.5}$ and $Y_{0.95}Ca_{0.05}BaCo_2O_{5.5}$ were prepared by standard solid state reaction technique. The precursors used were $Y_2O_3$, $BaCO_3$, $CaCO_3$ and $Co_3O_4$. In a first step, a stoichiometric mixture of the required precursors was intimately ground and heated at 900°C in air for 12 hrs for decarbonation. The mixture was then reacted in air at 1000°C for 24 hrs and slowly cooled to room temperature. After a regrinding of the powder, the annealing at 1000°C was repeated for another 24 hrs. This process was repeated several times, and after each annealing, the sample was slowly cooled to room temperature. Before the final annealing, the powder was pressed in the form of rectangular bars. The phase purity of the sample was checked using X-ray diffraction (XRD). No impurity phases were detected. The X-ray diffraction patterns were registered with a Panalytical X'Pert Pro diffractometer with a Co source (λ = 1.79 Å) under a continuous scanning mode in the 2θ range 5° - 150° and step size Δ2θ = 0.017°. The oxygen content of the samples was determined by iodometric titration. The

d.c. magnetization measurements were performed using a superconducting quantum interference device (SQUID) magnetometer with variable temperature cryostat (Quantum Design, San Diego, USA). All the magnetic properties were registered on dense ceramic bars of dimension ~ $4 \times 2 \times 2$ mm$^3$.

**Results and discussion**

*Structural Characterization*

The X-ray diffraction patterns (Fig. 1) confirm the phase purity of our samples, which stabilize in the orthorhombic symmetry with the *Pmmm* space group. The Rietveld analysis from the XRD data was done using the FULLPROF refinement program [29]. The fits are also shown in Fig. 1 (black curves). The bottom blue curves correspond to the difference between the observed and the calculated diffraction patterns. Satisfactory matching of the experimental with the calculated profile of the XRD pattern and the corresponding reliability factors (shown in Fig. 1) confirm that the fits obtained are reasonably accurate. The extracted lattice parameters for the two phases are also shown in Fig. 1. The oxygen stoichiometry for both samples (YBa$_{0.95}$Ca$_{0.05}$Co$_2$O$_{5.5+\delta}$ and Y$_{0.95}$Ca$_{0.05}$BaCo$_2$O$_{5.5+\delta}$) was fixed at $\delta = 0.00$ (1) from iodometric titration.

*D. C. magnetization study*

*Standard Zero Field Cooling and Field Cooling measurements*

The magnetization vs temperature curves for YBa$_{0.95}$Ca$_{0.05}$Co$_2$O$_{5.5}$ (Fig. 2(a)) and Y$_{0.95}$Ca$_{0.05}$BaCo$_2$O$_{5.5}$ (Fig. 2(b)) in the Zero Field Cooled (ZFC), Field Cooled Cooling (FCC) and Field Cooled Warming (FCW) modes recorded under a magnetizing field of H = 1 T are quite similar. In the ZFC mode, the samples were cooled from the paramagnetic state (T = 350 K) to the lowest measured temperature (10 K) in zero field, then the field was switched on and the measurement was made while warming up the samples. In the FCC mode, the applied magnetic field was switched on at T = 350 K, and the measurement was made while cooling the samples across the transition temperature to the lowest measured *T*. After completion of measurement in the FCC mode, the data points were again recorded in the presence of the same applied field while warming up the sample. This constituted the FCW mode. A fixed rate of temperature variation of 1 K/min was used throughout the study.

Both samples exhibit a paramagnetic to ferromagnetic phase transition near room temperature followed by a ferromagnetic to antiferromagnetic transition at lower temperature. These transitions are similar to those seen in the undoped sample (i.e. without Ca doping) [30]. However, as stated before, our zone of interest lies in the low temperature region, and we focus on the values of the magnetization achieved via the different measurement protocols at T = 10 K (the lowest measured temperature). A marked thermomagnetic irreversibility is seen in the samples, which increases with decreasing temperature. This thermomagnetic irreversibility is a direct consequence of phase coexistence in these samples which was reported earlier in Ref. 21. Aurelio et. al. [21] have reported that 10 % Ca doping at the Ba site in $YBaCo_2O_{5.5}$ suppresses the antiferromagnetic transition and induces ferrimagnetic order at low temperature. However, for 5 % Ca doping [21], the original antiferromagnetic order (that was present in the undoped phase) coexists with the ferrimagnetic order. The magnetization values obtained at low temperature will then depend on the relative phase fraction of the two phases. This is why the magnetization value of these samples is greater than that of the undoped phase (which is purely antiferromagnetic) at low temperature [20 - 21]. From our data (Fig. 2), we see that this relative phase fraction varies, depending on the measurement protocol. We obtain distinctly higher magnetization values for the FCC / FCW curves compared to the ZFC curve at low temperature. This is because via the field cooled measurement protocol, we are able to collect a greater volume fraction of the ferro (or ferri) magnetic phases than what was obtained via the zero field cooled measurement protocol, resulting in a higher magnetization value for the FC data.

This thermomagnetic irreversibility at T = 10 K can be quantified as the difference between the magnetization values obtained at T = 10 K via the ZFC and the FCC measurement protocols i.e., $\Delta M = M_{FCC\,(T\,=\,10\,K)} - M_{ZFC\,(T\,=\,10\,K)}$. The value of $\Delta M$ is seen to increase monotonically with an increase in the applied magnetic field (insets in Fig. 2). This is different from conventional magnetic systems, where $\Delta M$ actually decreases with an increase in the magnetic field. In our samples, the dependence of $\Delta M$ with H reflects the fact that we are able to progressively collect more and more volume fraction of the ferro (or ferri) magnetic phase by increasing the externally applied magnetic field. Thus, the coexisting ferro (or ferri) magnetic and antiferromagnetic phases in the sample can be easily tuned. This thermomagnetic irreversibility is essentially a result of the kinetics of the first order transition getting hindered, leading to a nonequilibrium magnetic state with a configuration of ferro (or ferri) magnetic and antiferromagnetic clusters frozen randomly at low temperature. This low temperature state is termed the *magnetic glass* state. The onset of glass transformation can be defined as the temperature ($T_g$) where the $M_{FCC}(T)$ curve starts flattening out (Fig. 2). In the next section we explore a more striking feature of this magnetic glass-like state.

*Cooling and Heating in Unequal Field (CHUF)*

In this section, we probe the magnetic glass state of $YBa_{0.95}Ca_{0.05}Co_2O_{5.5}$ and $Y_{0.95}Ca_{0.05}BaCo_2O_{5.5}$ using an experimental protocol which is in contrast to the standard FCC and FCW measurement protocols where the applied magnetic field while cooling and warming the sample is the same. In the current protocol, which has been termed as "Cooling and Heating in Unequal Field (CHUF)", the sample is cooled across the transition temperature in a certain applied magnetic field ($H_{cool}$). After the sample reaches the lowest temperature $H_{cool}$ is isothermally changed to a different field ($H_{measure}$) which may be larger or smaller than $H_{cool}$, and the magnetization is measured while warming the sample in the presence of this $H_{measure}$. The behaviour of our samples under the CHUF experimental protocol (Fig. 3) are very similar. The specially designed CHUF measurement protocol serves to bring out a special feature of the magnetic glass state as will be explained below.

The curves shown in Fig. 3 can be easily differentiated into two groups depending on their low temperature behaviour. For the curves for which $H_{cool} \leq H_{measure}$ (curves 1 – 4 in Fig. 3 (a) and (b)), the M(T) curves consist of only one sharp structure (marked by the black arrow in Fig. 3 (a) and (b)). This corresponds to the sharp rise in the magnetization signalling the antiferromagnetic to ferromagnetic transition in the sample. In contrast, for $H_{cool} > H_{measure}$ (curves 5 and 6 in Fig. 3 (a) and (b)), we get two sharp structures (marked by the two pink arrows in Fig. 3 (a) and (b)). These two sharp structures can be explained as follows: With a higher value of $H_{cool}$, the state obtained at the lowest temperature has a larger fraction of the kinetically arrested ferro (or ferri) magnetic component in the magnetic glass state. When the sample is subsequently warmed up, this glasslike arrested ferro (or ferri) magnetic phase fraction devitrifies and the system tries to approach the equilibrium antiferromagnetic phase, resulting in the rapid decrease in the magnetization. The second sharp structure corresponds to the same antiferromagnetic to ferromagnetic transition in the sample, as in the earlier case.

The CHUF experimental protocol, clearly showing the devitrification of the arrested state, thus, gives an unambiguous and rather visual evidence of the coexisting phases in the magnetic glass state.

*Time dependence of magnetization*

In this section, we probe the time dependence of the low temperature magnetization of $YBa_{0.95}Ca_{0.05}Co_2O_{5.5}$ and $Y_{0.95}Ca_{0.05}BaCo_2O_{5.5}$, which serves to explore the characteristic metastable behaviour associated with this sample. Thus, we have recorded the magnetization of the sample at T = 20 K as a function of time, both in the ZFC as well as the FC states. In the ZFC mode, the samples were cooled from T = 350 K to T = 20 K in zero field, they were then held at T = 20 K for a time $t_w$ = 300 sec, after which a magnetic field of H = 2 T was applied and the magnetization was recorded as a function of time. In the FC mode, the samples were cooled from T = 350 K to T = 20 K in the presence of an applied magnetic field of H = 2 T, they were then held at T = 20 K for a time $t_w$ = 300 sec, after which the magnetic field was removed and the magnetization was recorded as a function of time. The results obtained following the above mentioned protocols for the two samples are shown in Fig. 4. The magnetization values have been normalized with respect to the intial magnetization obtained at time t = 0.

It is seen that the systems show strong relaxation in magnetization in both the ZFC as well as the FC states. This is in contrast to other magnetic glass systems like $Ce(Fe_{0.96}Ru_{0.04})_2$ [31] and $La_{0.5}Ca_{0.5}MnO_3$ (see inset of Fig. 4) which show no relaxation of M in the ZFC state, the ZFC state being an equilibrium antiferromagnetic state in such cases. For our samples, it is clear that neither the ZFC state nor the FC state are completely in equilibrium. The reason for this can be traced back to the reported coexistence of antiferromagnetic and ferro (or ferri) magnetic phase fractions in these samples [21, 22]. However, the FC state shows more relaxation, as is evident from Fig. 4. The change in magnetization after the ZFC process is only ~ 2 % and ~ 4 % for $YBa_{0.95}Ca_{0.05}Co_2O_{5.5}$ and $Y_{0.95}Ca_{0.05}BaCo_2O_{5.5}$ respectively, while for the same time interval, the change in magnetization after the FC process in the two samples is almost 8 % and 12 % respectively.

Further, in order to study the evolution of the arrested glassy state, we measured magnetization as a function of time at different temperatures. The results for both samples are shown in Fig. 5. For measurement at each temperature, the samples were cooled in a magnetic field of H = 1 T from T = 350 K to T = 10 K. The magnetic field was then isothermally reduced to zero at T = 10 K, following which, the samples were warmed to the measurement temperature and magnetization was measured as a function of time for 100 min. Fig. 5 shows the evolution of M after it is normalized to the respective values of M at t = 0. In the figure, the curves are labelled with the respective temperatures at which they were recorded in the same color code for clarity.

When the samples are cooled from T = 350 K in a magnetic field of H = 1 T, they develop a frozen glassy ferro (or ferri) magnetic phase fraction. This glassy fraction remains invariant when the magnetic field is isothermally reduced to zero at T = 10 K. Magnetization decreases more rapidly with time as the measurement temperature is increased from T = 75 K to T = 120 K (see the curves recorded at T = 75 K, 85 K, 100 K and 120 K in Fig. 5 (a) and (b)). This observation is consistent with the kinetics being faster at higher temperature. However, the time evolution of magnetization starts decreasing for temperatures above T = 120 K (see the curves recorded at T = 130 K, 140 K and 150 K in Fig. 5 (a) and (b)). This indicates that above T = 120 K, the ferro (or ferri) magnetic glass fraction is left in a local minimum of free energy, and excitation over the free energy barrier is required to convert it to the antiferromagnetic phase. The barrier rises sharply as the temperature is raised above T = 120 K, as can be seen from the curves taken at T = 140 K and T = 150 K. This increased barrier causes a slower conversion even though the thermal energy is higher.

**Concluding remarks**

The formation of a magnetic glass-like state at low temperature in the "112" family of cobaltites has been shown for the first time in the present study of calcium doped $YBaCo_2O_{5.5}$ cobaltites. This state arises from the kinetic arrest of a first order ferromagnetic to antiferromagnetic phase transition in the sample, and consists of a glasslike ferro (or ferri) magnetic phase coexisting with the antiferromagnetic phase at low temperature. The influence of a second thermodynamic variable, the magnetic field, is also shown clearly in this temperature induced phenomenon. The magnetic glass state in Ca-doped $YBaCo_2O_{5.5}$ is seen even when the sample is cooled in zero field. This makes it different from some other magnetic glass systems like the alloy $Ce(Fe_{0.96}Ru_{0.4})_2$, where the glassy state is not present in the zero field cooled state, but develops only above a certain critical cooling field. The glasslike phase can also be destroyed by the process of devitrification. We have also shown that the ratio of the coexisting ferro (or ferri) magnetic and antiferromagnetic phases at low temperature in our sample can be easily controlled or tuned. This tunability provides a control over its functional properties.

Finally, the issue of the role of calcium for inducing such a phenomenon has to be understood. It is somewhat puzzling that the magnetic properties do not depend on the site of Ca substitution, either Y or Ba site. Indeed, substitution of Ca on the Y site is expected to increase the Co valency, in contrast to substitution at the Ba site, and should therefore, modify the magnetization differently. However, as the two samples behave almost identically, the only way to explain this unambiguous feature is to consider that Ca doping induces some very non-intuitive changes in the sample. Previous studies of the "112" $Ln_{1-x}Ca_xBaCo_2O_{5.5}$ cobaltites [22] have also shown that the expansion of ferromagnetism at the cost of antiferromagnetism could not be explained by a simple increase of the cobalt valency alone, but rather by a local disproportionation of $Co^{3+}$ into $Co^{4+}$ and $Co^{2+}$. Thus, both investigations converge to propose that cobalt disproportionation is the predominant effect. As a consequence, Ca doping induces local phase separation in the form of ferro or ferrimagnetic $(Co^{4+})_n$ clusters, in competition with antiferromagnetic $(Co^{2+})_n$ clusters, leading to a "$Co^{4+}/Co^{2+}$" magnetic glass. Such a disproportionation mechanism thus explains the fact that the properties of this magnetic glass do not depend on the location of calcium, which may sit either in Ba or in Y sites.

**Acknowledgements**

The authors acknowledge the CNRS and the Conseil Regional of Basse Normandie for financial support in the frame of Emergence Program. V. P. acknowledges support by the ANR-09-JCJC-0017-01 (Ref: JC09_442369).

**Figure Captions**

**Fig. 1** X-ray diffraction pattern along with the fit for (a) $YBa_{0.95}Ca_{0.05}Co_2O_{5.5}$ and (b) $Y_{0.95}Ca_{0.05}BaCo_2O_{5.5}$.

**Fig. 2** Magnetization vs temperature curves for (a) $YBa_{0.95}Ca_{0.05}Co_2O_{5.5}$ and (b) $Y_{0.95}Ca_{0.05}BaCo_2O_{5.5}$ following the ZFC, FCC and FCW measurement protocols under a magnetic field of H = 1 T. The insets show the variation of $\Delta M$ as a function of magnetic field.

**Fig. 3** Magnetization vs temperature for (a) $YBa_{0.95}Ca_{0.05}Co_2O_{5.5}$ and (b) $Y_{0.95}Ca_{0.05}BaCo_2O_{5.5}$ measured using the CHUF experimental protocol (see text for details).

**Fig. 4** Magnetization vs time at T = 20 K for $YBa_{0.95}Ca_{0.05}Co_2O_{5.5}$ and $Y_{0.95}Ca_{0.05}BaCo_2O_{5.5}$ obtained under ZFC and FC protocols. The inset shows the time dependence of $La_{0.5}Ca_{0.5}MnO_3$ under similar ZFC and FC protocols.

**Fig.5** Magnetization vs time for (a) $YBa_{0.95}Ca_{0.05}Co_2O_{5.5}$ and (b) $Y_{0.95}Ca_{0.05}BaCo_2O_{5.5}$ at different temperatures.

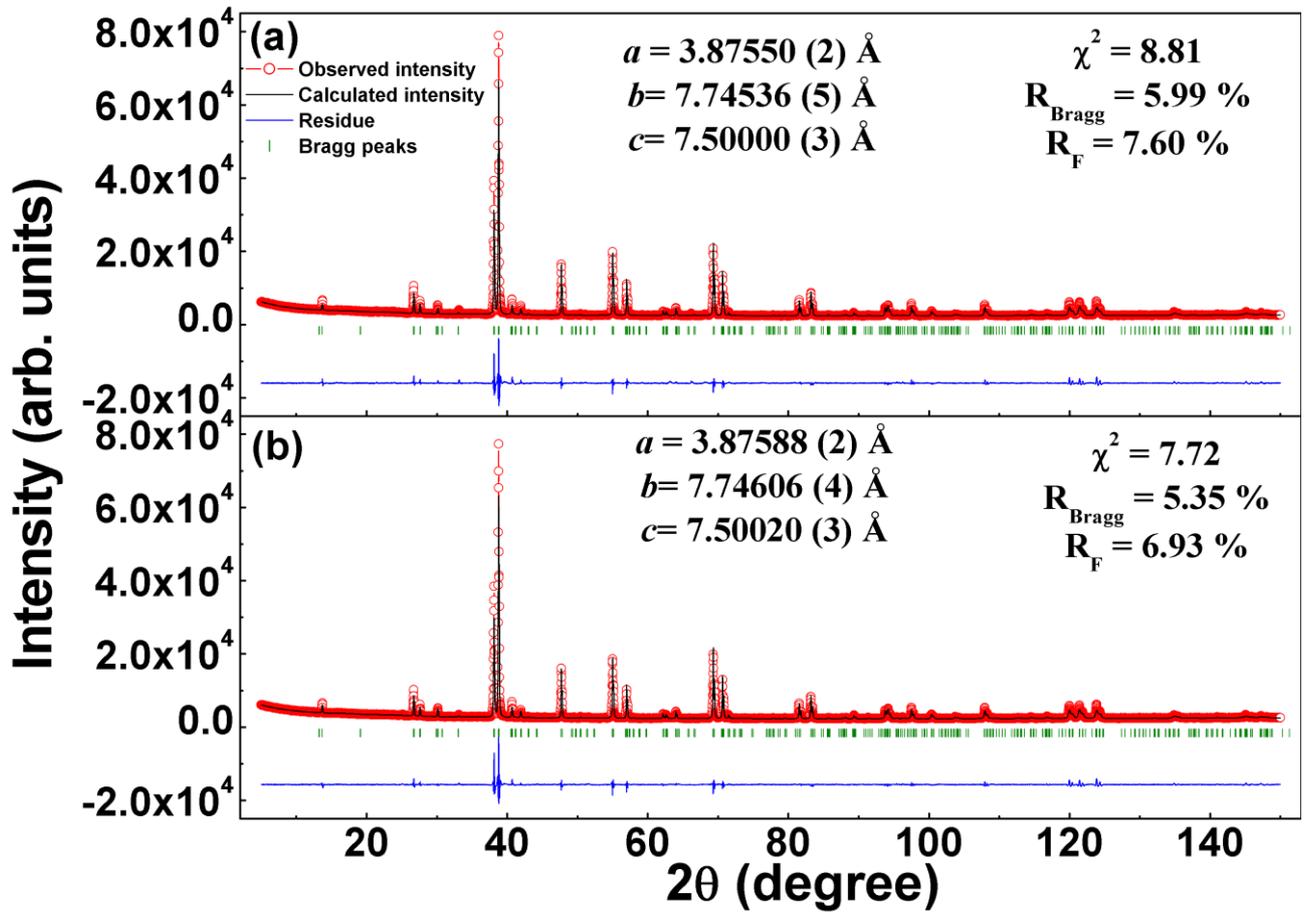

Fig. 1.

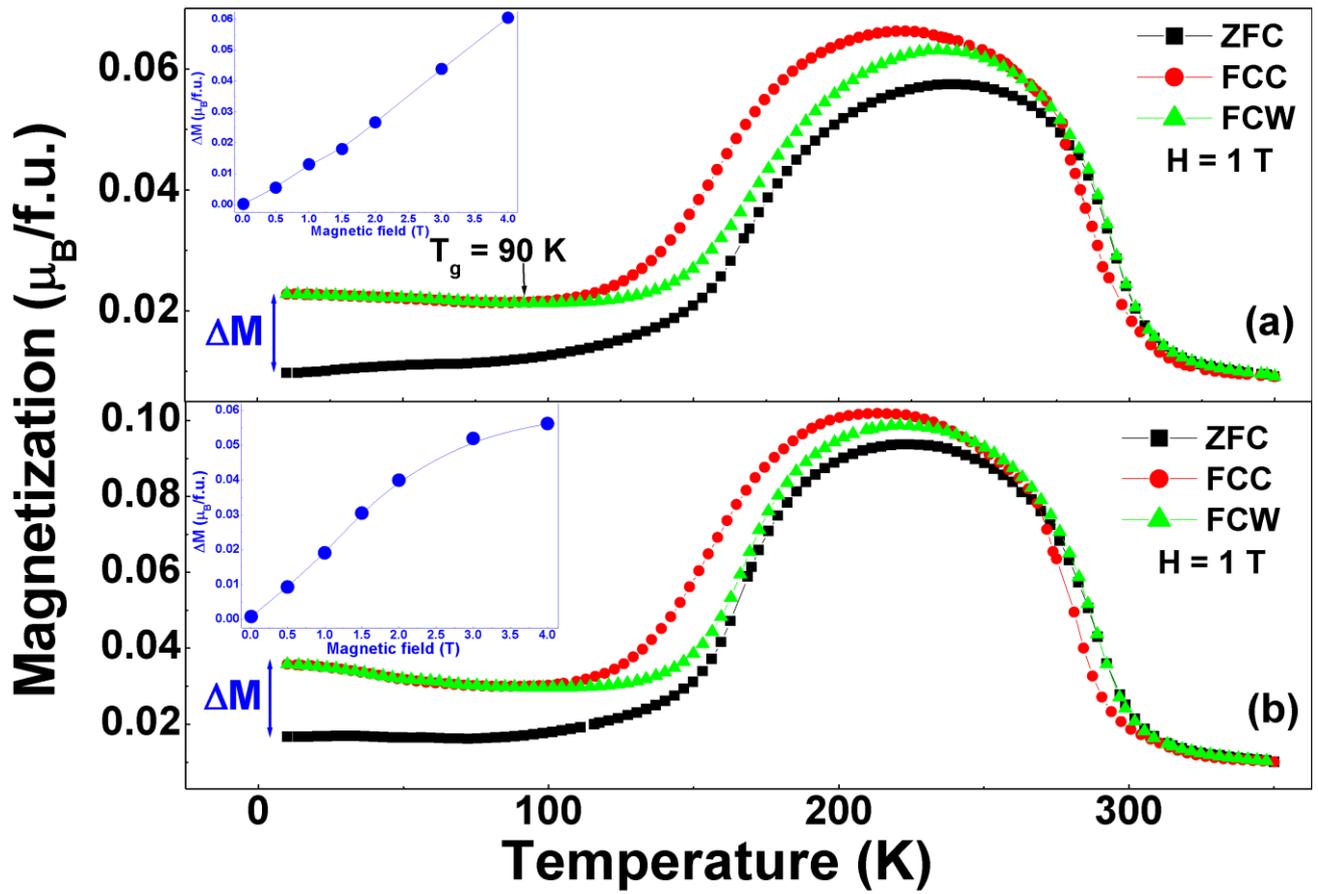

Fig. 2.

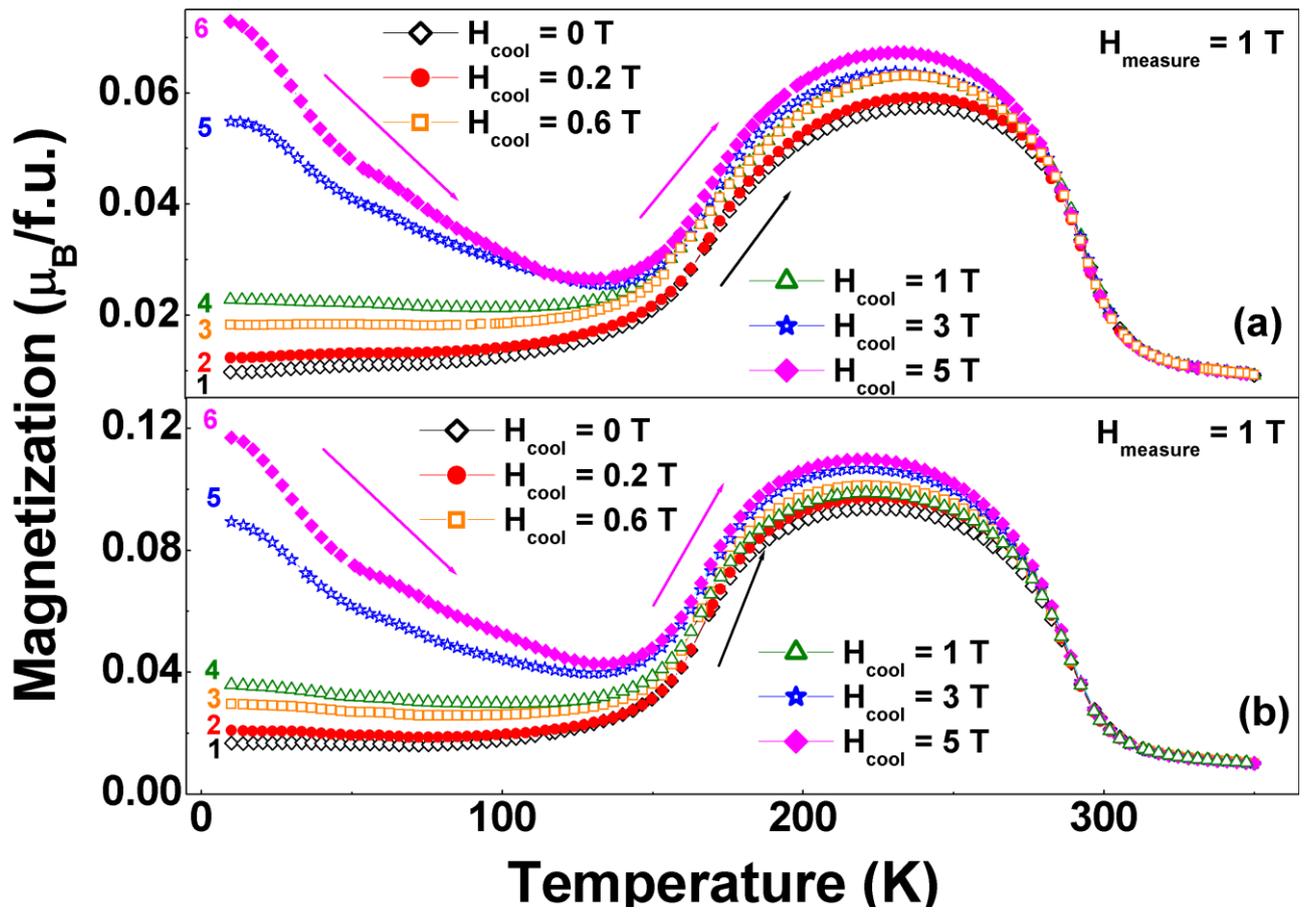

Fig. 3.

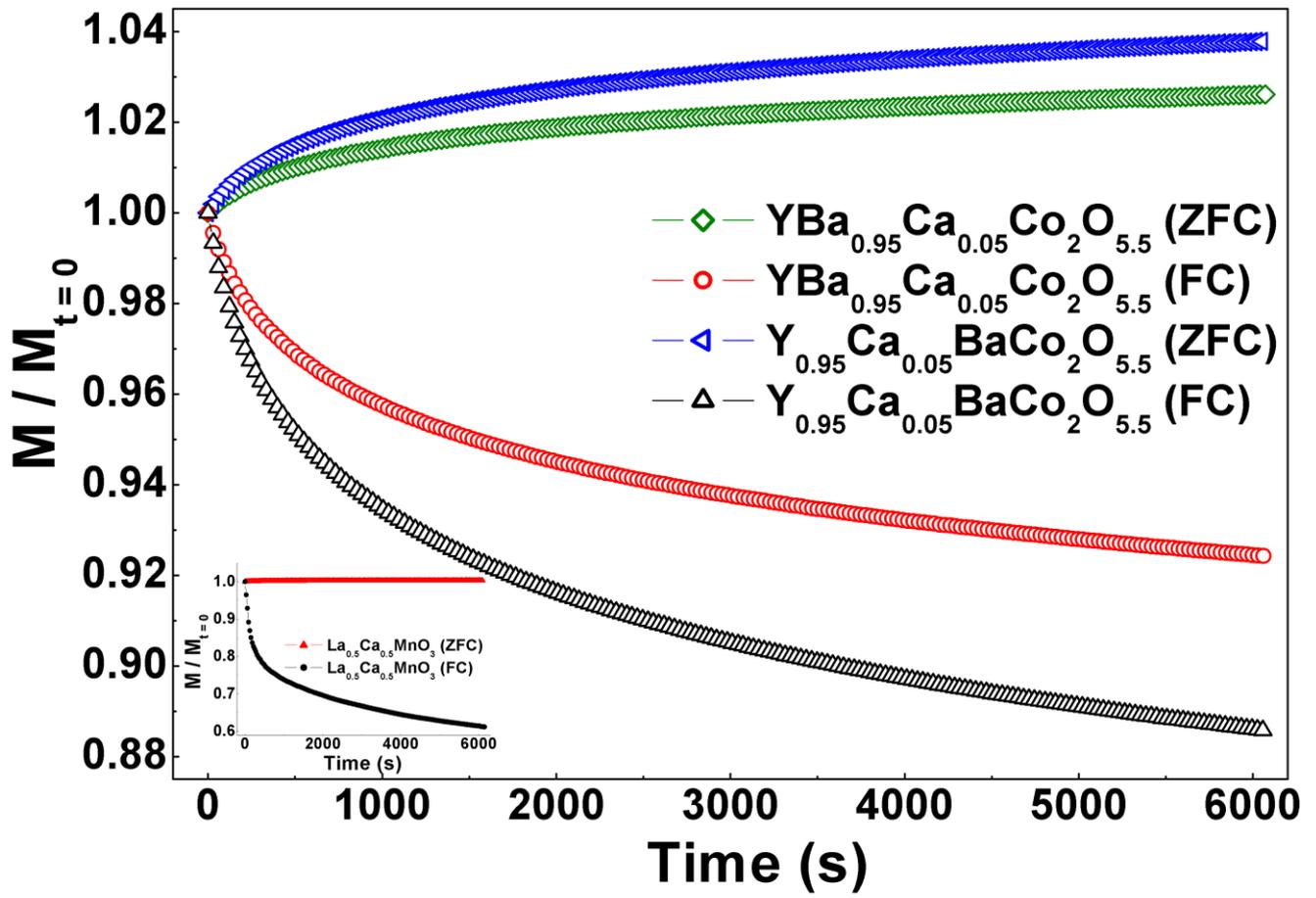

Fig. 4.

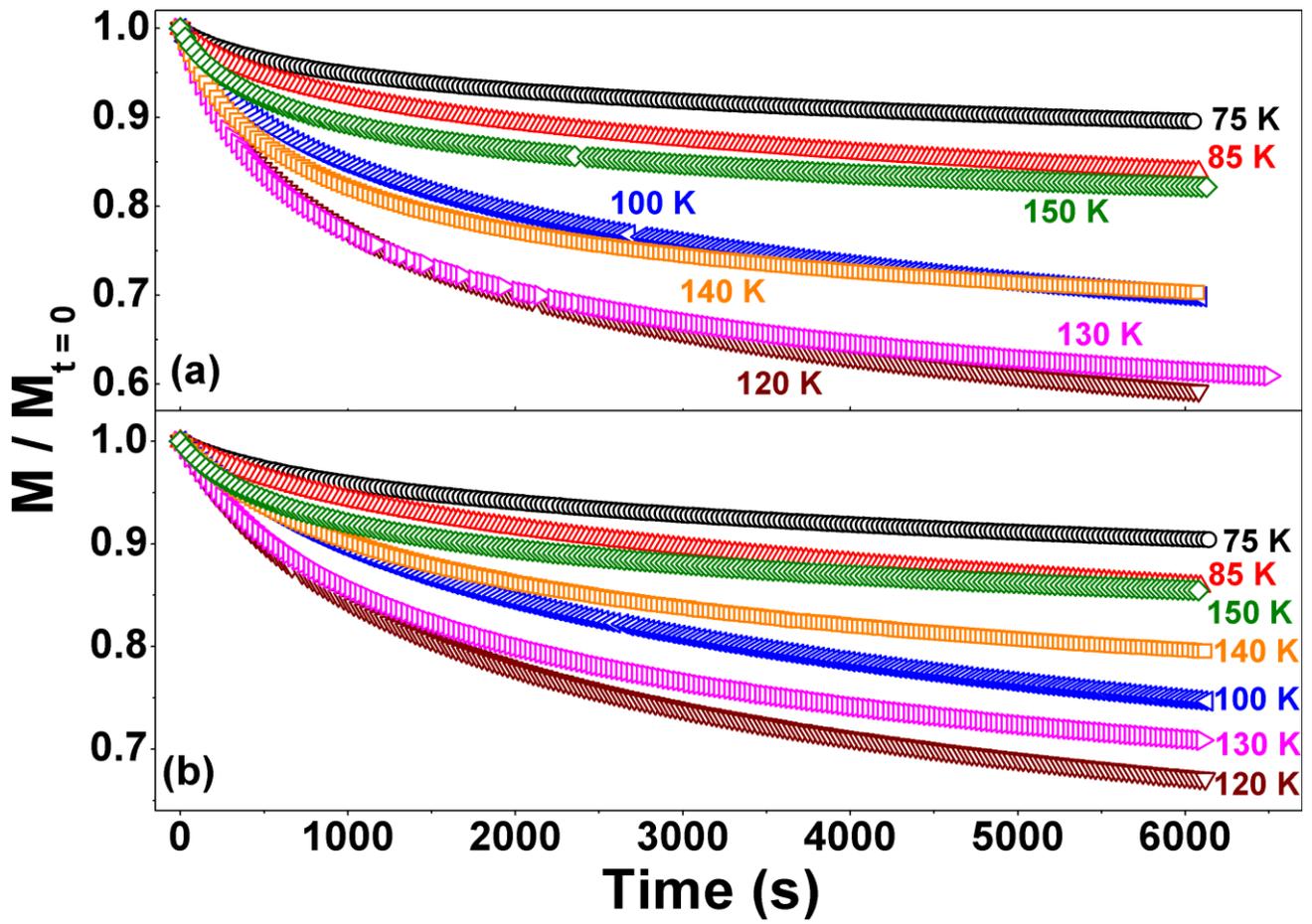

Fig. 5.